\documentclass[journal=nalefd,layout=twocolumn]{achemso}
\pdfoutput=1
\usepackage{amssymb}
\usepackage{amsmath}
\usepackage{graphicx}
\usepackage{setspace}
\usepackage{epstopdf}
\usepackage{color}
\newcommand\RED[1]{\textcolor{black}{#1}}

\newcommand{\GO}{${\rm{G_1}}$}
\newcommand{\GT}{${\rm{G_2}}$}

\newcommand{\RO}{${\rm{S_1}}$}
\newcommand{\RT}{${\rm{S_2}}$}
\newcommand{\RJ}{${\rm{S}_j}$}

\newcommand{\Vds}{$V_{\rm{ds}}$}

\title{Level spectrum and charge relaxation in a silicon double quantum dot probed by dual-gate reflectometry}
\author{Alessandro Crippa}
\email{alessandro.crippa@cea.fr}
\affiliation{Universit\'e Grenoble Alpes \& CEA INAC-PHELIQS, F-38000 Grenoble, France}
\alsoaffiliation{Dipartimento di Scienza dei Materiali, Universit\`a di Milano Bicocca, Via Cozzi 53, 20125 Milano, Italy}
\alsoaffiliation{CNR-IMM, Via C. Olivetti 2, 20864 Agrate Brianza (MB), Italy}

\author{Romain Maurand}
\affiliation{Universit\'e Grenoble Alpes \& CEA INAC-PHELIQS, F-38000 Grenoble, France}

\author{Dharmraj Kotekar-Patil}
\affiliation{Universit\'e Grenoble Alpes \& CEA INAC-PHELIQS, F-38000 Grenoble, France}

\author{Andrea Corna}
\affiliation{Universit\'e Grenoble Alpes \& CEA INAC-PHELIQS, F-38000 Grenoble, France}

\author{Heorhii Bohuslavskyi}
\affiliation{Universit\'e Grenoble Alpes \& CEA INAC-PHELIQS, F-38000 Grenoble, France}
\alsoaffiliation{CEA, LETI MINATEC campus, F-38000 Grenoble, France}

\author{Alexei O. Orlov}
\affiliation{Department of Electrical Engineering, University of Notre Dame, Notre Dame, Indiana 46556, USA}

\author{Patrick Fay}
\affiliation{Department of Electrical Engineering, University of Notre Dame, Notre Dame, Indiana 46556, USA}

\author{Romain Lavi\'eville}
\affiliation{Universit\'e Grenoble Alpes \& CEA INAC-PHELIQS, F-38000 Grenoble, France}
\alsoaffiliation{CEA, LETI MINATEC campus, F-38000 Grenoble, France}

\author{Sylvain Barraud}
\affiliation{Universit\'e Grenoble Alpes \& CEA INAC-PHELIQS, F-38000 Grenoble, France}
\alsoaffiliation{CEA, LETI MINATEC campus, F-38000 Grenoble, France}

\author{Maud Vinet}
\affiliation{Universit\'e Grenoble Alpes \& CEA INAC-PHELIQS, F-38000 Grenoble, France}
\alsoaffiliation{CEA, LETI MINATEC campus, F-38000 Grenoble, France}

\author{Marc Sanquer}
\affiliation{Universit\'e Grenoble Alpes \& CEA INAC-PHELIQS, F-38000 Grenoble, France}

\author{Silvano De Franceschi}
\affiliation{Universit\'e Grenoble Alpes \& CEA INAC-PHELIQS, F-38000 Grenoble, France}

\author{Xavier Jehl}
\affiliation{Universit\'e Grenoble Alpes \& CEA INAC-PHELIQS, F-38000 Grenoble, France}

\begin{document}

\begin{abstract}
We report on dual-gate reflectometry in a metal-oxide-semiconductor double-gate silicon transistor operating at low temperature as a double quantum dot device. The reflectometry setup consists of two radio-frequency resonators respectively connected to the two gate electrodes. By simultaneously measuring their dispersive responses, we obtain the complete charge stability diagram of the device. Electron transitions between the two quantum dots and between each quantum dot and either the source or the drain contact are detected through phase shifts in the reflected radio-frequency signals. At finite bias, reflectometry allows probing charge transitions to excited quantum-dot states, thereby enabling direct access to the energy level spectra of the quantum dots. Interestingly, we find that in the presence of electron transport across the two dots the reflectometry signatures of interdot transitions display a dip-peak structure containing quantitative information on the charge relaxation rates in the double quantum dot.\\
\end{abstract}

\maketitle

Keywords: dispersive readout, reflectometry, double quantum dot, charge relaxation, high-frequency resonator, silicon.
\\

\begin{figure}
 \centering
 \includegraphics[width=\columnwidth]{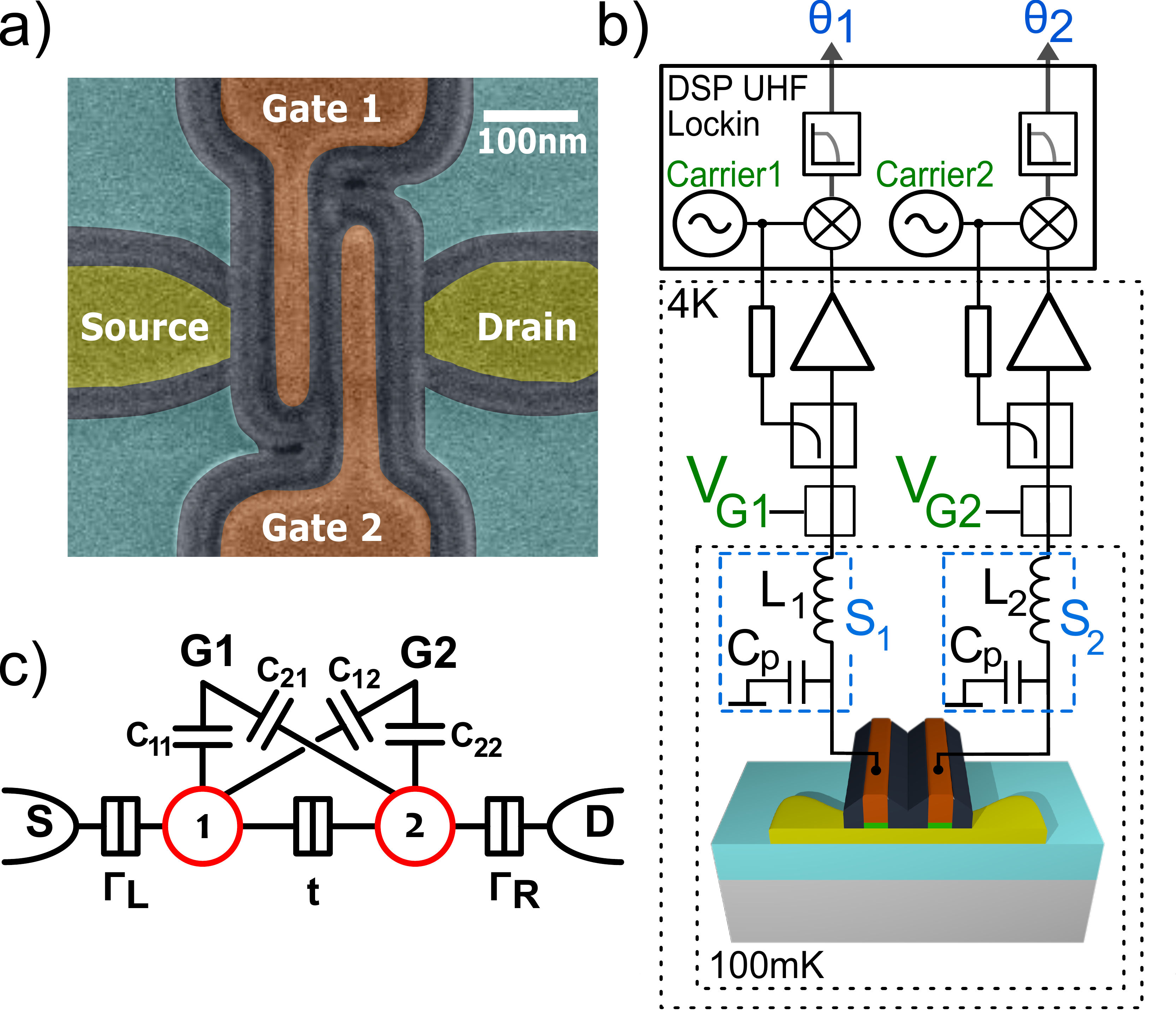}
 \caption{a) False color scanning electron micrograph of a typical double gate device nominally identical to the measured samples. Orange denotes the gate electrodes, whereas spacers are dark grey. Heavily-doped regions of reservoirs are highlighted in yellow. b) Schematic of the dual-port measurement setup. Each gate is connected to a resonator for dispersive readout. Demodulation is performed at room temperature by a Ultra High Frequency dual lock-in. c) Equivalent circuitry of the sensed quantum system. The two quantum dots are red circles. Labels and elements are explained in the main text.}
\label{fig:setup}
\end{figure}

\begin{figure}
 \centering
 \includegraphics[width=\columnwidth]{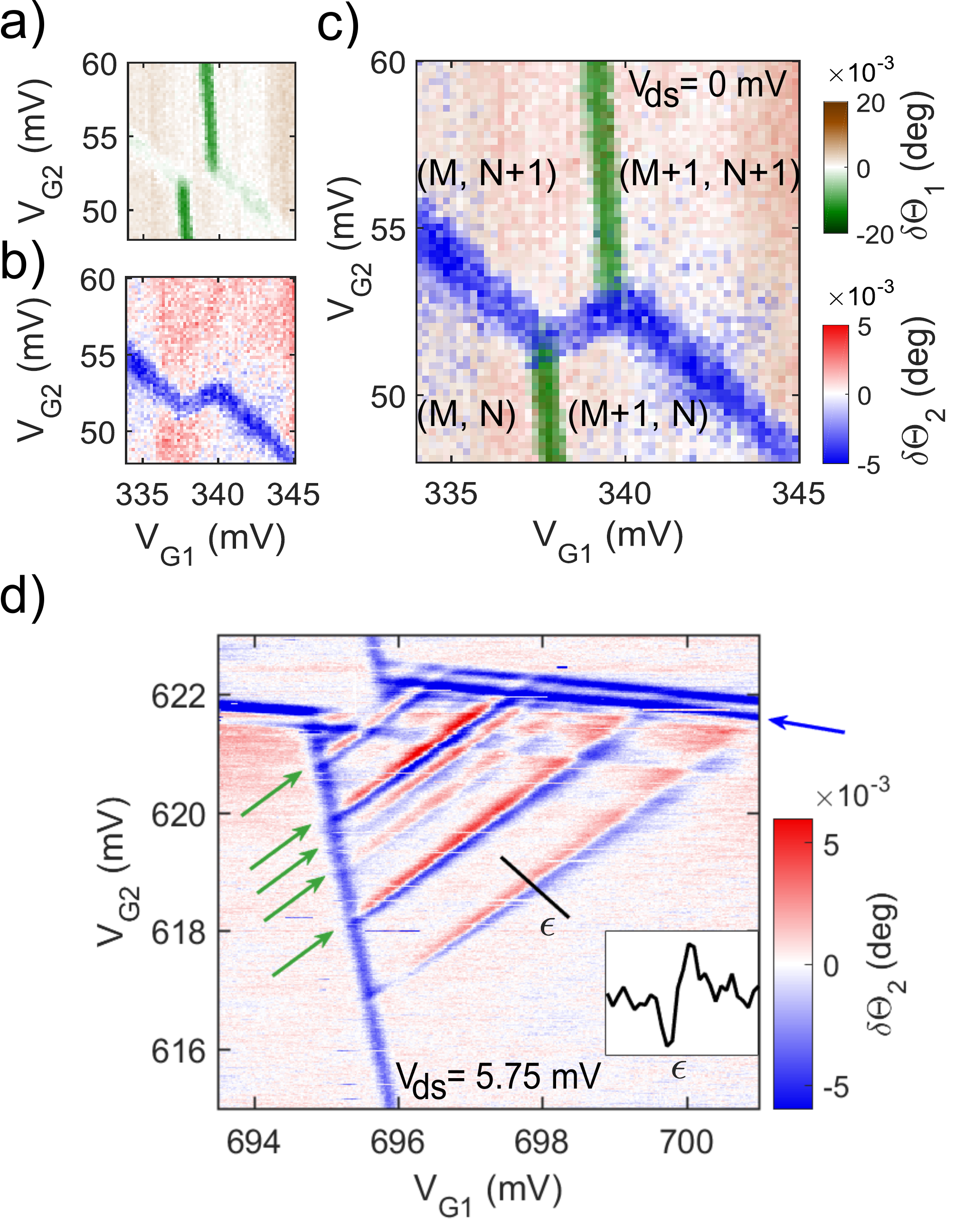}
 \caption{\RED{a), b) Charge stability diagrams around triple points at nominal 0 bias from detector S1 and S2 respectively. c) Overlap of panels a) and b). Here $M \lesssim 5, N\lesssim 10 $. d) Triple points of another sample with $M, N \sim 15$, measured at $V_{ds} = 5.75$ mV}. The plot in d) exhibits the orbital spectrum of the DQD through lines parallel to the base of the triangles and lines parallel to the edges.}
\label{fig:triangle}
\end{figure}

\begin{figure*}[h]
 \centering
 \includegraphics[width=2\columnwidth]{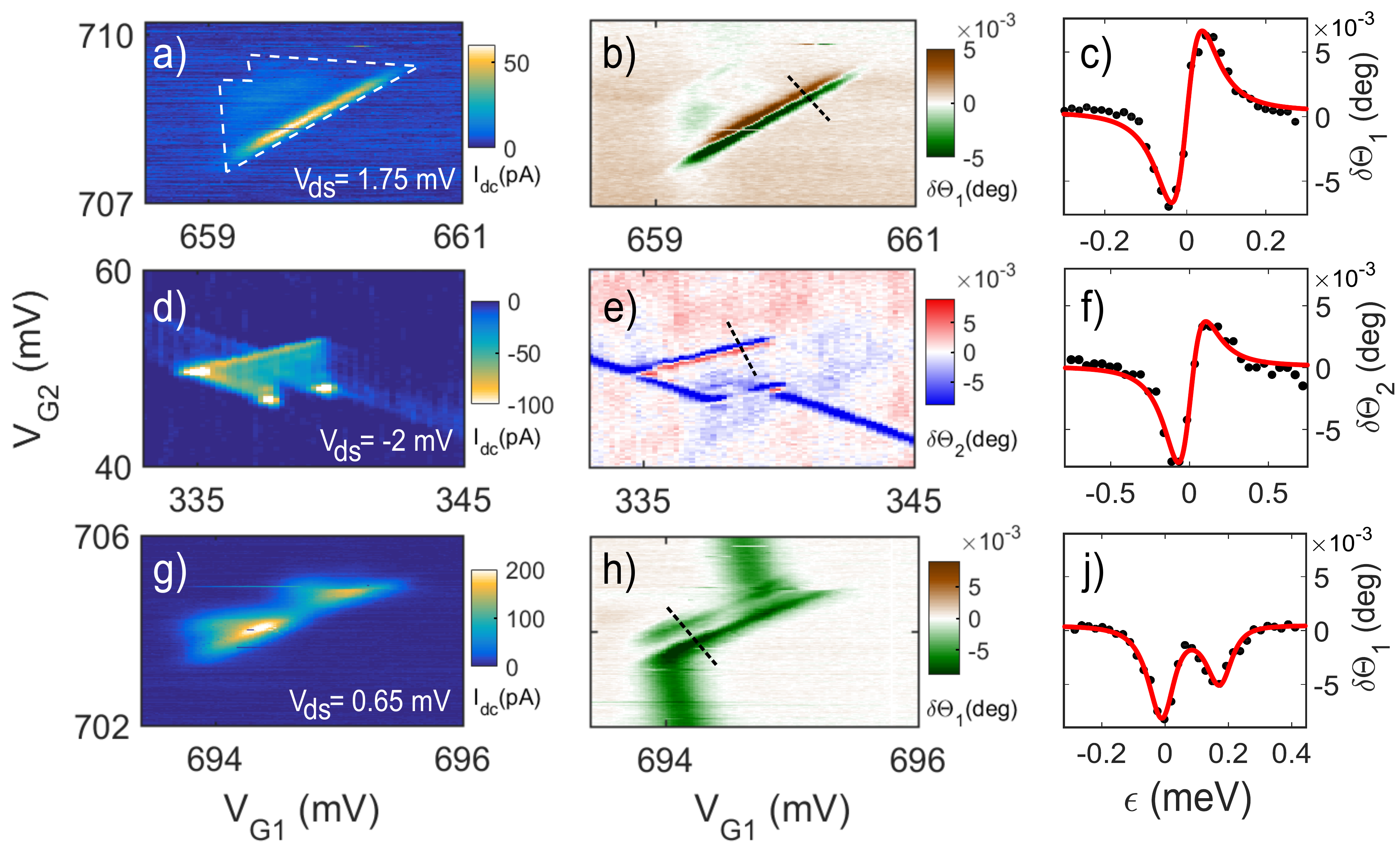}
 \caption{Stability diagrams of bias triangles recorded by source-drain current [panels a), d), f)] and dispersive signal [panels b), e), h)] relative to three different triple points characterized respectively by weak, medium and fast relaxation rate $\gamma$. Reflectometry data are plotted for the gate detector with better interdot signal. Black dashed lines indicate the detuning axis setting the energy scale on the plots for fitting on the right [c), f) and j)]. Triple point edges in a) are marked by white dashed lines as a guide for the eyes.}
\label{fig:comparison}
\end{figure*}

Integration of charge sensors for the readout of quantum bits (qubits) is one of the necessary ingredients for the realization of scalable semiconductor quantum computers \cite{Reilly_2015npjqi}.
In most qubits developed so far in silicon, like electron or nuclear spins in quantum dots and single atoms \cite{pla_electronqubit, Vandersypen_electronqubit, pla_nuclearqubit}, charge \cite{Erikkson_chargequbit} or hybrid spin-charge states \cite{Erikkson_hybrid}, qubit readout has been performed with the aid of quantum point contacts (QPCs) or single-electron transistors (SETs). These charge-sensitive devices, however, involve a significant overhead in terms of gates and contact leads, posing an issue for scalability towards many-qubit architectures.\\ 
Gate-coupled radio-frequency (RF) reflectometry \cite{Ciccarelli, Reilly_2013PRLdispersive, Betz_limit} has been recently proposed as an alternative approach to qubit readout. In this technique, the charge sensing required to sense the qubit state is accomplished by measuring the dispersive response of an electromagnetic RF resonator connected to one of the qubit gates and excited at its resonance frequency. 
The absence of local charge sensors simplifies the qubit physical layer \cite{Reilly_2015npjqi} and allows for a tighter qubit pitch. Further, the RF resonator can be a microscopic circuit shared by many qubits through frequency multiplexing \cite{Reilly2014_multiplexing}.\\
Before its first implementation in a gate-coupled geometry \cite{Ciccarelli}, RF reflectometry has been (and still is) applied to increase the readout bandwidth and sensitivity of electrometers such as QPCs, SETs or Cooper-pair transistors \cite{Dzurak07_RFQPC, Hirayama2002_APL, Delsing04_JAP, Delsing_CQdef}. Also, it has been used to discern remote charge traps in the polysilicon gate stack \cite{Villis_2011APL} or in the channel \cite{Villis_2014APL} of silicon nanowire field-effect transistors (FETs).\\
Gate-coupled reflectometry has been shown \cite{Betz_limit} to achieve bandwidths and sensitivities comparable to those of on-chip charge detectors.  Combined with spin-to-charge conversion processes \cite{TaruchaPSB}, it has allowed detecting spin-related effects in double quantum dots (DQDs) \cite{Petta_wireAPL, Petta_parity, Petersson_2010NL, Betz_SB2015NL}.
In a recent experiment, gate reflectometry has also been used to reveal coherent charge oscillations in a silicon DQD \cite{FernandoLZS} and 
measure coherence times of the order of 100 ps, consistent with those found in an earlier study based on DC transport \cite{LZSsilvano}.\\
Here we apply gate reflectometry to investigate the electronic properties of a silicon DQD. Besides probing the full charge stability diagram at equilibrium (i.e. zero drain-source bias voltage $V_{ds}$), we shall present reflectometry measurements at finite $V_{ds}$, where we will show that i) gate reflectometry allows probing excited quantum dot levels (even when the DC current is too small to be measured), and ii) the dispersive response of a resonator is strongly influenced by the conductive regime of the DQD and the associated charge relaxation rate.\\
Measurements are carried out on double-gate silicon nanowire FETs fabricated on a 300-mm, silicon-on-insulator (SOI) wafer using an industry-standard complementary metal oxide semiconductor (CMOS) processing facility. Device fabrication relies entirely on deep ultra-violet lithography except for one step, based on electron-beam lithography and used in the definition of the two gates, whose spacing is well below optical resolution \cite{Roro_holequbit}.
Fig.\,\ref{fig:setup}a) shows a false color scanning electron micrograph of a typical device. Transistors present a silicon channel thickness $t_{\text{Si}}=11$\,nm and width $W_{\text{Si}}=15$\,nm, whereas gates, denoted as \GO{} and \GT{}, are patterned with a length $L_g=35$\,nm and a spacing $S_{gg}=35\,$nm. At low temperature, the two gates are employed in accumulation mode to define two electron quantum dots in series.\\
DC charge transport measurements are performed by grounding the source electrode, S, and connecting the drain contact, D, to a transimpedance amplifier with adjustable reference potential. 
Two lumped-element resonators, \RO{} and \RT{}, are connected to \GO{} and \GT{}, respectively. Each resonator is composed of a surface-mount inductor (nominal inductances: $L_1=270$\,nH and $L_2=390$\,nH for \RO{} and \RT{}, respectively) and the parasitic capacitance $C_p$ at the corresponding gate \cite{Ciccarelli, Reilly_2013PRLdispersive, Betz_limit, Alexei_dualport} [see Fig.\,\ref{fig:setup}b)]. The resulting resonance frequencies are $f_0^{(1)}= 421$ MHz and $f_0^{(2)}=335$ MHz, respectively. Unlike drain-based reflectometry, where the LC circuit provides impedance-matching conditions for the source-drain resistance of the charge sensor (SET or QPC), here the purpose of the inductor along with its parasitic capacitance is to provide a resonant network sensitive to small changes in the load capacitance (in this case the total capacitance seen by the gate). The dispersive responses of \RO{} and \RT{} are simultaneously recorded using homodyne detection on the reflected RF signals.
\\A circuit representation of the DQD system is shown in Fig.\,\ref{fig:setup}c). We label the quantum dot accumulated below \GO{} (\GT{}) as ``dot 1'' (``dot 2'').
The electrostatic influence of gate $j$ on dot $i$ ($i,j=1,2$) is mediated by a capacitance $C_{ij}$. 
The DQD is electrically connected to the S and D reservoirs via tunnel barriers with characteristic tunnel rates $\Gamma_{\rm{L}}$ and $\Gamma_{\rm{R}}$, respectively. 
We operate the device in a regime where $k_BT_e \lesssim 2t < \Delta$, with $k_B$ is the Boltzmann constant, \RED{$T_e \sim 300$ mK is the electron temperature}, $t$ the interdot tunnel coupling and \RED{$\Delta$ is the mean level spacing in each dot [typical range in our devices is from $\sim0.1$ to few meV \cite{Kondo_Crippa, Dharam_PSB}, consistently with Figs.\,\ref{fig:triangle}d) and \ref{fig:comparison}h)]}.
Resonator \RO{} (\RT{}) is sensitive through \GO{} (\GT{}) to changes in the quantum admittance of the DQD system \cite{Cottet}. More precisely, an electron tunneling between two dots or between a dot and a reservoir gives rise to a small capacitance $C^j_{\text{diff}}$ seen by the resonator \RJ{} in addition to the geometric capacitances of Fig.\,\ref{fig:setup}c). The differential capacitance of a dot coupled to gate $j$ is 
\begin{equation}
\label{eq:diffCap}
C^{j}_{\text{diff}}=-e \alpha_{j} \frac{\partial \langle \nu \rangle}{\partial V_{Gj}}.
\end{equation}
$e$ is the electron charge, $\alpha_{j}$ is the lever-arm to convert the gate voltage $V_{Gj}$ into energy and $\langle \nu \rangle$ is the average excess charge flowing through the dot.\\ 
If the charge dynamics is faster than the probing frequency $f_0^{(j)}$ of resonator \RJ{}, the reflected signal experiences a phase variation $\delta \Theta_j \propto -C^j_{\text{diff}}$  \cite{Delsing_CQdef}, hence non zero in correspondance of the charge transitions.\\
Let's now define the formalism for an isolated DQD. No charges are given to or removed from the reservoirs but a single electron can be switched between the two dots, corresponding to an excess electron in either dot 1 or dot 2. Interdot dynamics between the localized states $|1\rangle$ and $|2\rangle$, respectively of dot 1 and dot 2, is described by the Hamiltonian $H=(\epsilon/2) \sigma_z + t \sigma_x$: $\epsilon$ is the detuning parameter, i.e. the misalignement between chemical potentials of the two dots in the limit of vanishing interdot coupling $t$, and $\sigma_{z,x}$ are Pauli matrices. The eigenstates of such two-level system are
\begin{equation}
\label{eq:psi}
\begin{split}
& |\psi_- \rangle = \sin(\theta/2) |1\rangle - \cos(\theta/2)|2\rangle \\
& |\psi_+ \rangle = \cos(\theta/2) |1\rangle + \sin(\theta/2)|2\rangle,
\end{split}
\end{equation}
where $\tan \theta = 2t/\epsilon$ and $|\psi_-\rangle$, $|\psi_+\rangle$ correspond to the bonding and antibonding molecular states expressed in terms of the localized states $|1\rangle$ and $|2\rangle$.
The two eigenvalues, basically the states of a charge qubit, are $E_{\pm}=\pm \sqrt{\epsilon^2 + (2t)^2} /2$. For small excitations applied by gate $j$ to dot $i$, i.e. $\delta V_{Gj}\ll 2t/e\alpha_{ij}$ (see Supporting Information), the differential capacitance for resonator \RJ{} reads \cite{FernandoLZS}
\begin{equation}
\label{eq:capacitances}
C_{\text{diff}}^j= \frac{\beta_j^2}{2} \Bigl( \frac{dP_{\text{d}}}{d\epsilon} \frac{\epsilon}{\sqrt{\epsilon^2+(2t)^2}} +P_{\text{d}} \frac{(2t)^2}{[\epsilon^2+(2t)^2]^{3/2}} \Bigr),
\end{equation}
with $P_{\text{d}} = P_- - P_+$ the difference between the occupation probabilities of $|\psi_-\rangle$ and $|\psi_+ \rangle$; $\beta_j \equiv -e(\alpha_{1j} - \alpha_{2j})$ is the detuning lever-arm factor for gate $j$. The first term of Eq. (3), called tunnel capacitance, counts for transitions whose occupation probabilities $P_-$ and $P_+$ depend on the detuning, whereas the second one, named quantum capacitance, is related to band curvature $\partial^2 E_{\pm}/\partial \epsilon^2$ \cite{Delsing_CQdef}.\\
We now present our experimental results. Fig.\,\ref{fig:triangle} contains a representative set of stability diagrams measured by gate reflectometry. Panels a) and b) display color maps of the phase responses $\delta \Theta_1$ and $\delta \Theta_2$ obtained from \RO{} and \RT{}, respectively. These two data sets have been simultaneously recorded for the same range of gate voltages V$_{\rm{G1}}$ and V$_{\rm{G2}}$ at $V_{ds}=0$.
Denoted the electron number in dot 1 and dot 2 as $M$ and $N$ respectively, dot-lead charge transitions $(M,N) \leftrightarrow (M+1,N)$ ($(M,N+1) \leftrightarrow (M,N)$) are detected primarily by resonator S1 (S2). Interdot transitions $(M+1,N) \leftrightarrow (M,N+1)$ in principle could be equally sensed by both S1 and S2; in the case of Fig.\,\ref{fig:triangle}, however, they happen to be clearly visible only in the $\delta \Theta_2$ phase plot. Interdot lines can also be found in $\delta \Theta_1$ signal in other regions of the stability diagram (not shown).
Such behavior is probably related to the fact that the two quantum dots are not precisely positioned under the respective gates, and that their exact location changes with the electron filling of the channel. 
Panel c) is simply the superposition of a) and b). It clearly shows that the two RF resonators together allow detecting all charge boundaries in the stability diagram. \RED{In the rest of the paper we will always show the most representative dispersive signal between $\delta \Theta_1$ and $\delta \Theta_2$.}\\
Let us now focus on interdot charge transitions. Since $\,hf_0^{(j)} < 2t$ ($h$ is Planck's constant), interdot tunneling is adiabatic and $C_{\text{diff}}$ is dominated by the quantum capacitance associated to $|\psi_-\rangle$, leading to a negative phase shift, i.e. a dip in the dispersive signal \cite{Betz_SB2015NL}. This result is expected at equilibrium, where the DQD lies in its ground state $|\psi_-\rangle$. Out of equilibrium, a non negligible probability of occupying the excited state $ |\psi_+ \rangle$, which gives a positive phase shift, can qualitatively change the dispersive response. 
For instance, Ref.\,\citenum{FernandoLZS} reports a strongly driven two-level system (i.e. $hf>2t$) in which non-adiabatic transitions between $ |\psi_- \rangle$ and $ |\psi_+ \rangle$ result in a Landau-Zener-St$\rm{\ddot{u}}$ckelberg interference pattern with both negative and positive phase shifts in the reflected signal.
Indeed Eq. (3) holds also for DQDs out of equilibrium as long as the resonators sense the stationary occupation propabilities of the DQD states. 
In the following, we investigate the out-of-equilibrium regime resulting from a finite \Vds{}. At positive detuning the excited state $|\psi_+ \rangle$ is populated in the process of charge tunneling from the source reservoir. 
Fig.\,\ref{fig:triangle}d) shows the phase response on \RT{} in a conducting regime of the DQD at \Vds{}$=5.75$ mV. The first thing we learn from this measurement is that, similarly to current transport, gate dispersive readout can detect excited-state transitions. These transitions, denoted by green arrows, form a set of lines running parallel to the base of the triangular regions, which corresponds to ground-state tunnel events \cite{ReviewDQD}.
Given the sign of the applied bias voltage, electrons flow from dot 1 to dot 2. As a result, the observed lines correspond to excited states of dot 2. On the other hand, the line denoted by a blue arrow, running parallel to the almost horizontal edge of the upper triangle is associated with an excitation in dot 1 occurring during tunneling from the source contact to dot 1.\\
We remark that the demonstrated use of gate reflectometry as a tool for energy-level spectroscopy could be extrapolated to DQDs with extremely weak coupling to the source and drain reservoirs, where the more standard transport-based spectroscopy is hindered by unmeasurably small current levels.\\ 
Fig.\,\ref{fig:triangle}d) bares a second important message. All inter-dot transitions display a clear peak-dip structure. This is revealed by a measurement of the phase response as a function of the detuning $\epsilon$, see the $\delta \Theta_2(\epsilon)$ trace in the inset to Fig.\,\ref{fig:triangle}d) corresponding to a line cut along the black solid line.
\begin{figure}
 \centering
 \includegraphics[width=\columnwidth]{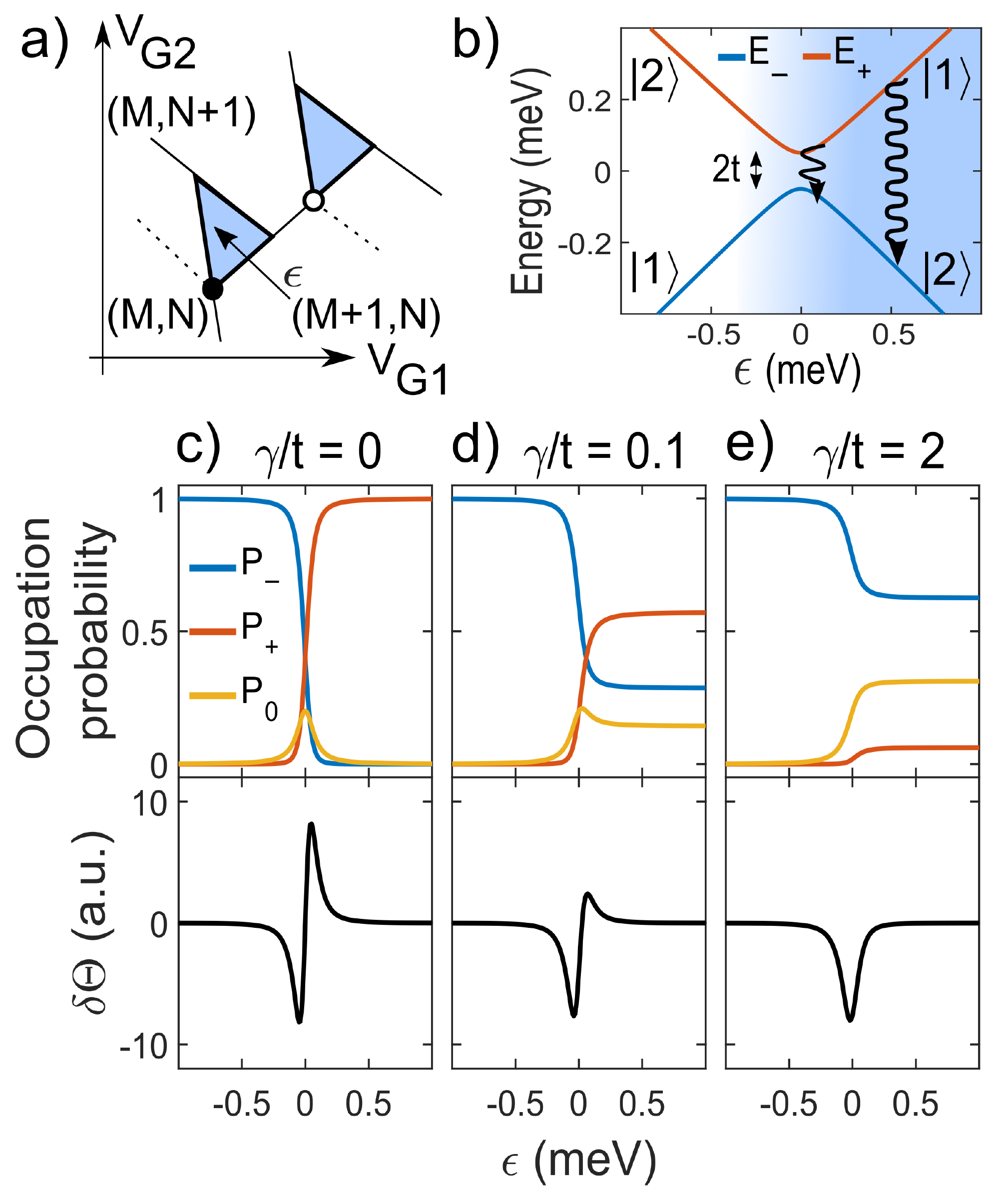}
 \caption{Schematic stability diagram around two triple points at small bias voltage [contrary to Fig.\,\ref{fig:triangle}d), here transport through excited states is not visible]. a) Schematic stability diagram at finite bias voltage. Numbers in the brackets denote the fixed-charge domains of the DQD; the light-blue triangular areas highlight the conductance regions. b) Ground and excited states of the DQD in presence of an extra charge. The light-blue background indicates the conductive condition. In the hybridization region $|\epsilon|<2t$ the $E_{\pm}$ band curvature is maximized: resonators are sensitive to elastic and inelastic charge transfers. In the far detuned limit $\epsilon > 2t$, $|\psi_-\rangle$ and $|\psi_+\rangle$ are more localized, hence well approximated by single dot states $|1\rangle$ and $|2\rangle$: inelastic transitions are weakly sensed due to the lack of band curvature. c), d), e) Occupation probabilities of the ground ($P_-$), excited ($P_+$) and empty ($P_0$) configuration of the DQD from Eq. (4). The correspondant phase shift $\delta \Theta$ is computed from Eq. (3). When there are no inelastic interdot transitions [$\gamma/t=0$, panel c)], the dispersive response is a symmetrical dip-peak line shape. Such resonance gradually becomes asymmetric by setting some charge relaxations [$\gamma/t=0.1$, panel d)], till it is just a dip for fast relaxations [$\gamma/t=2$, panel e)]. In these plots $\Gamma_R=10 \, \mu$eV, $\Gamma_L=20 \, \mu$eV and $t=50 \, \mu$eV.}
\label{fig:theory}
\end{figure}
\RED{To understand the physical origin of the observed peak-dip structures we focus on the charge transport regime involving the ground states of both dots, corresponding to a regime around $\epsilon=0$}. We initially derive a set of master equations for the occupation probabilities of the DQD states. 
At positive $V_{ds}$, electron transport involves cyclic transitions in the charge states of the DQD, notably $(M+1,N) \to (M,N+1) \to (M,N) \to (M+1,N)$ 
for the lower triangular region of Fig.\,\ref{fig:theory}a), and ($(M+1,N) \to (M,N+1) \to (M+1,N+1) \to (M+1,N)$) for the upper one. For the sake of simplicity, let us consider only the first of these two options. These corresponding charge states can be mapped onto the ``molecular'' basis states $\{|0\rangle, |\psi_-\rangle, |\psi_+\rangle \}$,  where $|0\rangle$ refers to the $(M,N)$ charge configuration. 
The rate equations for the corresponding occupation probabilities are \cite{Ensslin_MW}:
\begin{equation}
\label{eq:master}
\begin{split}
& \dot{P_0}=-\Gamma_L P_0+ \zeta \Gamma_R P_-+ \Gamma_R(1-\zeta) P_+\\
& \dot{P_-}=\Gamma_L(1-\zeta) P_0 - \zeta \Gamma_R P_- + \gamma P_+ \\
& \dot{P_+}=\zeta \Gamma_L P_0 + [\Gamma_R (\zeta-1) - \gamma] P_+
\end{split}
\end{equation}
with $P_0+P_-+P_+=1$, $\zeta \equiv \cos^2(\theta/2)$ and $\gamma$ the inelastic interdot tunnel rate [in other words, the relaxation rate between the energy levels $E_{\pm}$ of the charge qubit, see Fig.\,\ref{fig:theory}b)].
The stationary occupation probabilities $P_0, P_-, P_+$ depend on the dot-lead rates $\Gamma_L$, $\Gamma_R$ and the elastic and inelastic interdot rates, respectively $t$ and $\gamma$. \RED{In particular, $\gamma$ has to be seen as an average rate over the sensitivity region of the resonators around $\epsilon=0$.} Once estimated $P_+$ and $P_-$ from Eq. (4), the resulting phase response $\delta \Theta$ is computed by Eq. (3).\\
To perform the simulations of Figs.\,\ref{fig:theory}c), d) and e) we assume from the device symmetry the access barriers $\Gamma_L$, $\Gamma_R$ differ from one other by one order of magnitude at maximum; in particular, we notice that $\Gamma_L$ does not significantly influence the $\delta \Theta$ lineshape, whereas $\Gamma_R$ rules the height of the whole phase resonance. The excited state $|\psi_+\rangle$ originates the peak side of the dip-peak structure when $\Gamma_R>\gamma$. Finally, the full width at half maximum (FWHM) of the dip is found to be $\sim 2t$ when dip and peak are both present and $\sim 2.5 t$ for a dip-only resonance. Under these costraints, the asymmetry between the dip and peak amplitudes is set mostly by $\gamma/t$, as displayed in Figs.\,\ref{fig:theory}c), d) and e). 
In absence of energy relaxation ($\gamma = 0$, panel c)), charge transport makes $|\psi_-\rangle$ mainly occupied for $\epsilon<0$ and  $|\psi_+\rangle$  mainly occupied for $\epsilon>0$. Equal occupation probabilities are found for $\epsilon=0$. Since $E_-(\epsilon)$ and $E_+(\epsilon)$ have exactly opposite curvatures, the dispersive response $\delta \Theta(\epsilon)$ is characterized by a dip-peak structure with equally strong peak and dip components, as shown in the bottom panel of Fig.\,\ref{fig:theory}c). 
In presence of energy relaxation ($\gamma > 0$), the average probability to be at energy $E_+$ for positive detuning decreases, resulting in a reduced amplitude of the positive $\delta \Theta$ component (see bottom panel of Fig.\,\ref{fig:theory}d)). Finally, for a strong energy relaxation rate ($\gamma \gg t$), the DQD is predominantly in the $|\psi_-\rangle$ state for any detuning, and the dispersive response exhibits only a dip in $\delta \Theta$, as shown in Fig.\,\ref{fig:theory}e).
\\Fig.\,\ref{fig:comparison} provides an experimental demonstration of the three regimes of Figs.\,\ref{fig:theory}c), d) and e). \RED{Line cuts are taken from the dispersive maps away from the triple points to avoid thermal broadening effects from the leads, and then are fitted to Eq. (3).}
Panel a) shows a transport measurement with a pair of bias triangles in which the inelastic current is barely visible, indicating a slow relaxation rate. In the corresponding reflectometry signal, presented in panel b), the overlapping bases of the triangles appear with a pronounced dip-peak structure. Data along the dashed line are displayed in panel c) revealing that peak and dip have almost identical amplitudes, as expected for  
$\gamma/t \ll 1$. From the FWHM of the dip, $t=40\,\mu$eV. Given the assumptions above, the phase trace is fitted with fairly good agreement, yielding $\gamma=2 \pm 1\,\mu$eV, thus $\gamma/t \approx 1/20$.\\
The intermediate regime, where energy relaxation starts to be important, is observed on a different pair of bias triangles, where the inelastic current is small but clearly visible, as shown in panel d). A dip-peak structure is still visible in the corresponding reflectometry signal [see panel e) and line cut in panel d)], but the peak component is now weaker, as expected from a non negligible energy relaxation. For $t=90\,\mu$eV, the fit to Eq. (3) results again in a good agreement, yielding $\gamma/t= 0.08 \pm 0.02$.\\ 
The strong relaxation regime is observed on a third pair of bias triangles. The transport data in panel g) show a pronounced inelastic current. The (positive) peak component has completely disappeared from the phase response leaving only a dip structure. In fact, two dip structures are seen in panels h) and j). They correspond to different orbital levels of dot 2. Given $t=35\,\mu$eV, the fit gives $\gamma/t =3.0 \pm 0.3$, consistent with the observation of strong energy relaxation. \RED{Referring back to Fig.\,\ref{fig:triangle}d), we note that our fitting procedure is also applicable to interdot transitions to excited states, which potentially allows to correlate the dominant tunnel processes (elastic or inelastic) with the orbital character of the wave functions involved \cite{BurkardPetta}.}\\
\RED{To complete the discussion, the time scales to obtain coherent oscillations between the states of a DQD out of equilibrium can be found in Supplemental Material}.\\
In summary, we have studied silicon double quantum dots by means of a dual RF gate reflectometry setup. Two resonators operate in-parallel as independent charge transfer detectors. Such dispersive readout technique enables the mapping of DQD charge stability diagrams both at zero and finite bias voltage. For a net source-drain current across the DQD, the dispersive response is sensitive to charge relaxation between double-dot molecular states; by taking into account transport rate equations, we fit the reflected phase signal to extract the interdot tunnel coupling and the charge relaxation rate. 
Ref.\,\cite{ViennotPRB} and a recent proposal about Si/SiGe DQD dispersive readout by a superconducting microwave resonator \cite{BurkardPetta} testify the relevance of dispersive sensing of level structure and interdot coupling rates of DQDs, e.g. for metrology \cite{Jehl_PumpRelexation} and quantum information. Indeed, by expedients like frequency multiplexing \cite{Reilly2014_multiplexing, LairdExchangeOnly} and use of specific RF cryo-electronics \cite{HEMT_JAP}, multi-gate reflectometry may represent a viable technique for scalable single-shot readout of quantum dot arrays \cite{Jones_qdarray}.

\begin{suppinfo}
RF modulation amplitudes analysis and discussion on possible regimes to realize coherent oscillations.
\end{suppinfo}

\begin{acknowledgement}
The authors are grateful to M. L. V. Tagliaferri and M. Fanciulli for fruitful discussions. This work has been financially supported by the EU through the FP7 ICT collaborative projects SiAM (No. 610637) and SiSPIN (No. 323841).
\end{acknowledgement}

\bibliography{biblio}

\end{document}